\pdfoutput=1

\documentclass[
 twocolumn,
 reprint,superscriptaddress,
 amsmath,amssymb,
 aps,aip
]{revtex4-1}

\usepackage{docs}
\usepackage{bm}
\usepackage[usenames,dvipsnames]{color}
\usepackage{graphicx}
\usepackage{dcolumn}
\usepackage{hyperref}
\usepackage{float}
\usepackage[export]{adjustbox}
\usepackage{wrapfig}
\usepackage{amssymb}
\usepackage{blindtext}
\usepackage{tcolorbox}
\usepackage{tabularx}

\usepackage[left]{lineno}

\begin{document}

\title{A perspective on topological nanophotonics: current status and future challenges}

\author{Marie S. Rider}
\email{marie.rider16@imperial.ac.uk}
\affiliation{The Blackett Laboratory, Imperial College London, London SW7 2AZ, United Kingdom}

\author{Samuel J. Palmer}
\affiliation{The Blackett Laboratory, Imperial College London, London SW7 2AZ, United Kingdom}

\author{Simon R. Pocock}
\affiliation{The Blackett Laboratory, Imperial College London, London SW7 2AZ, United Kingdom}

\author{Xiaofei Xiao}
\affiliation{The Blackett Laboratory, Imperial College London, London SW7 2AZ, United Kingdom}

\author{Paloma Arroyo Huidobro}
\affiliation{The Blackett Laboratory, Imperial College London, London SW7 2AZ, United Kingdom}

\author{Vincenzo Giannini}
\homepage{www.GianniniLab.com}
\affiliation{Instituto de Estructura de la Materia (IEM-CSIC), Consejo Superior de Investigaciones Cient{\'i}ficas, Serrano 121, 28006 Madrid, Spain}
\affiliation{The Blackett Laboratory, Imperial College London, London SW7 2AZ, United Kingdom}


\begin{abstract}
Topological photonic systems, with their ability to host states protected against disorder and perturbation, allow us to do with photons what topological insulators do with electrons. Topological photonics can refer to electronic systems coupled with light or purely photonic setups. By shrinking these systems to the nanoscale, we can harness the enhanced sensitivity observed in nanoscale structures and combine this with the protection of the topological photonic states, allowing us to design photonic local density of states and to push towards one of the ultimate goals of modern science: the precise control of photons at the nanoscale. This is paramount for both nano-technological applications and also for fundamental research in light matter problems. For purely photonic systems, we work with bosonic rather than fermionic states, so the implementation of topology in these systems requires new paradigms. Trying to face these challenges has helped in the creation of the exciting new field of topological nanophotonics, with far-reaching applications. In this prospective article we review milestones in topological photonics and discuss how they can be built upon at the nanoscale. 
\end{abstract}

\maketitle

\section{Overview}
\label{sec:overview}
One of the ultimate goals of modern science is the precise control of photons at the nanoscale. Topological nanophotonics offers a promising path towards this aim. A key feature of topological condensed matter systems is the presence of topologically protected surface states immune to disorder and impurities. These unusual properties can be transferred to nanophotonic systems, allowing us to combine the high sensitivity of nanoscale systems with the robustness of topological states. We expect that this new field of topological nanophotonics will lead to a plethora of new applications and increased physical insight.

\begin{figure}[t]
\includegraphics[width=\columnwidth]{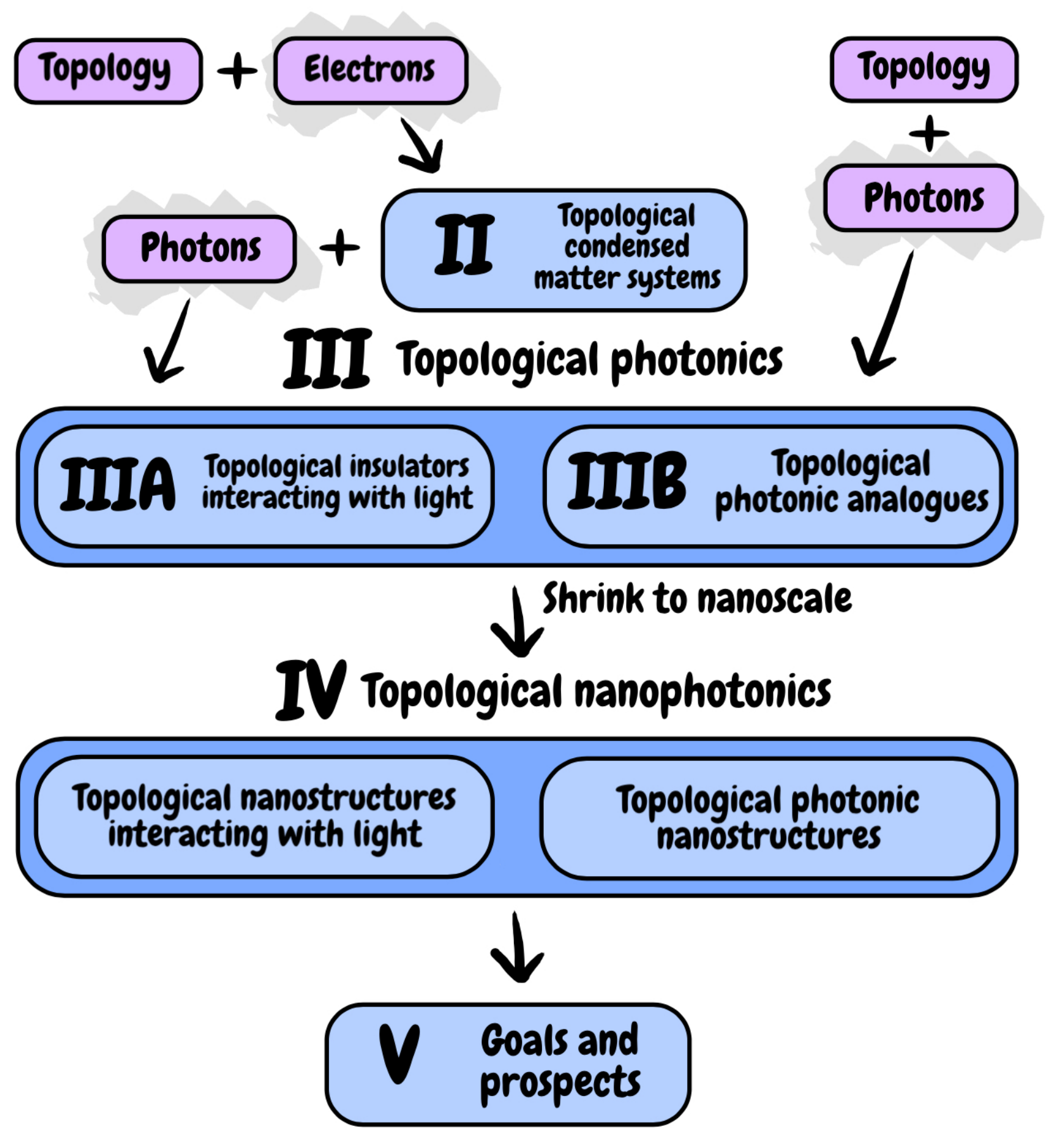}
\caption{\textbf{Schematic overview} Schematic showing the topics covered in this perspective. \label{fig:overview}}
\end{figure}

\begin{figure*}
\includegraphics[width=0.9\textwidth]{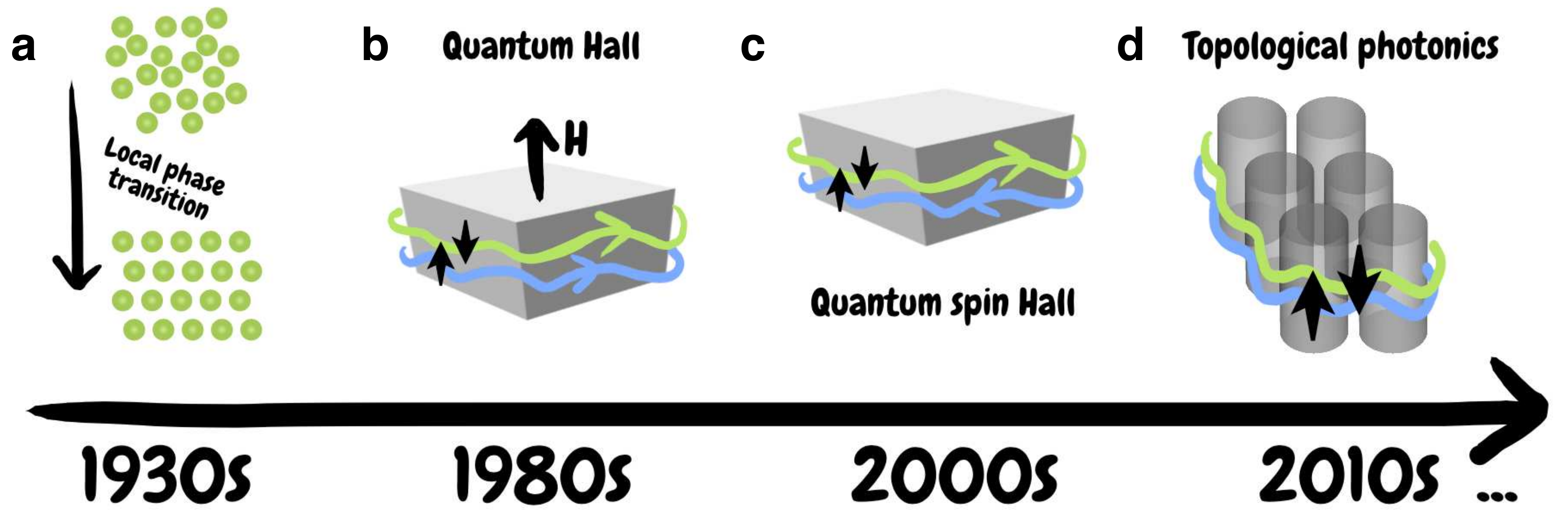}
\caption{\textbf{Topology in condensed matter systems} \textbf{(a)} Liquid-solid transition parameterised by a local order parameter. \textbf{(b)} Quantum Hall system, supporting topologically protected edge states. \textbf{(c)} Quantum spin Hall system (2D topological insulator). \textbf{(d)} Photonic topological insulator displaying edge states.  \label{fig:topo_history}} 
\end{figure*}

In this perspective, as presented schematically in FIG.~\ref{fig:overview}, we begin (section \ref{sec:topo_condensed}) by exploring topology in electronic systems. We aim this section towards readers who are new to the topic, so begin at an introductory level where no prior knowledge of topology is assumed. 

In section \ref{sec:topo_photonics} we introduce light, first by discussing how topological electronic systems can interact with light (section \ref{subsec:tis_with_light}), then move onto the topic of topological photonic analogues (section \ref{subsec:topo_photonic_analogues}), in which purely photonic platforms are used to mimic the physics of topological condensed matter systems. 

In section \ref{sec:topo_nano} we discuss various paths via which  topological photonics can be steered into the nanoscale. Excellent and extensive reviews already exist on topological photonics~\cite{ozawa2018topological,lu2014topological,khanikaev2017two,Sun2017}, and many platforms showcasing unique strengths and limitations are currently being studied in the drive towards new applications in topological photonics such as cold atoms~\cite{goldman2016topological}, liquid helium~\cite{volovik2003universe}, polaritons~\cite{PhysRevX.5.031001}, acoustic~\cite{fleury2015nonreciprocal} and mechanical systems~\cite{huber2016topological} but in this work we restrict ourselves to nanostructures. We discuss the efforts up until now in this very new field, and in section \ref{sec:goals} we review how the concept of topological photonic analogues can be built upon and surpassed in order to create photonic topological systems with no electronic counterpart, and we outline the open questions and challenges which must be faced and overcome in order to master topological nanophotonics.

\section{Topology in Condensed Matter Systems} 
\label{sec:topo_condensed}
Much of modern physics is built on the concept of symmetries and the resulting conserved quantities. We are most familiar with the symmetries and phases of matter characterised by local order parameters within the Landau theory of phase transitions~\cite{landau1980statistical, cardy1996scaling} (as in FIG. \ref{fig:topo_history}a), but in the last few decades the exploration of topological phases of matter has lead to many new developments in our understanding of condensed matter physics, culminating in a Nobel prize for Thouless, Haldane and Kosterlitz in 2016, and a Breakthrough prize in fundamental physics for Kane and Mele in 2019. The notion of topology in physics was introduced by von Klitzing and his discovery of the 2D quantum Hall (QH) state ~\cite{klitzing1980new}, with Thouless et al. explaining the quantization of the Hall conductance in 1982 ~\cite{thouless1982quantized}. Whereas QH states explicitly break time-reversal (TR) symmetry, new topologically non-trivial materials obeying TR symmetry have been discovered. The first proposals of the 2D topological insulator (TI) - otherwise know as the quantum spin Hall state (QSH) - were remarkably recent (Kane and Mele, 2005 ~\cite{kane2005z,kane2005quantum} and Bernevig and Zhang, 2006 ~\cite{bernevig2006quantum2}). The 3D generalisation came soon after in 2007 ~\cite{fu2007topological} and experiments have shown that these new phases of matter are both realisable and accessible ~\cite{konig2007quantum, hsieh2008topological, moore2010birth}. Other systems can have topology associated with another symmetry, such as the Su, Schrieffer, and Heeger (SSH) model~\cite{PhysRevLett.42.1698}, which owes its topological properties to sublattice symmetry.

\subsection{Topological invariants and band structures}
\label{subsec:topo_invariants}
Before the concept of topology was connected to condensed matter systems, phase transitions could only be characterized by local order parameters. For example, a disordered liquid, which when cooled will solidify to a crystal with long range order (as illustrated in FIG. \ref{fig:topo_history}a). A local order parameter such as mass density $\rho(\mathbf{r})$ can be defined, and constructed by looking only at a small neighbourhood around $\mathbf{r}$. A small deformation of the Hamiltonian may trigger $\rho(\mathbf{r})$ to grow discontinuously from 0, signalling a local phase transition.

\subsubsection{Topological phase transitions}

In contrast to theories of local phase transitions, topological phases of matter cannot be described with a local order parameter. Unlike the previously described system in which only the local neighbourhood around a point $\mathbf{r}$ contributed to the local order parameter, in a system exhibiting topology the whole system must be measured to ascertain the phase. For a system which can exist in multiple topological phases, no local order parameter can be constructed which will distinguish between the phases and we must instead rely on the idea of topological invariants. In condensed matter, a topological invariant is a global quantity which characterizes the Hamiltonian of the system, and a topological phase transition must occur to change the value of the invariant. Topologically trivial phases have a topological invariant equal to 0.

\begin{figure}[t]
\includegraphics[width=\columnwidth]{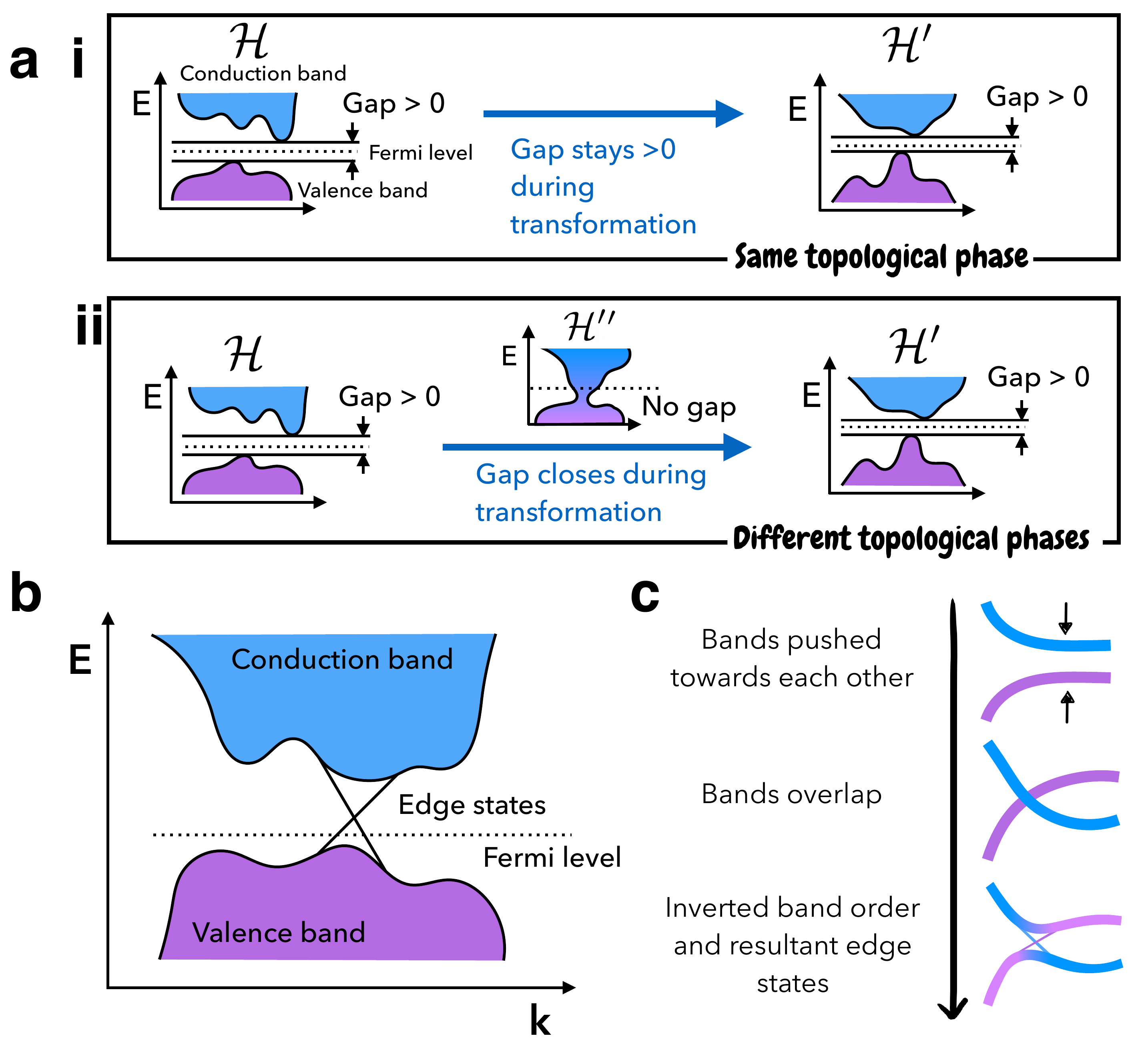}
\caption{\textbf{Topology and bandstructures} \textbf{(a)} \textbf{(i)} Gap remaining open during Hamiltonian transformation means the systems described by the two Hamiltonians are in the same topological phase, whilst \textbf{(ii)} a topological phase transition will result in the gap closing during the transformation.  \textbf{(b)} For a finite system comprising of a topologically non-trivial bulk surrounding by a trivial background, a topological phase transition occurs on the surface and the gap is closed. Gapless topological edge states exist on the surface, traversing the gap. \textbf{(c)} Band inversion due to mechanism such as spin-orbit coupling. \label{fig:topo_bands}}
\end{figure}

\subsubsection{Topological band structures and edge states}
For an insulating system described by the Hamiltonian $\mathcal{H}$, we may smoothly deform our system to one described by a new Hamiltonian $\mathcal{H}'$, as illustrated in FIG. \ref{fig:topo_bands}a. If the band gap remains open during the transformation (as shown in FIG. \ref{fig:topo_bands}a(i)), the number of states residing in the valence band is necessarily conserved, as although these states can mix amongst themselves during the transformation the only way for the number of states to increase is to close the gap and allow states to enter from or leave to the conduction band. The number of states in the valence band is a topological invariant, and will only change value if the band gap closes during a transformation (see shown in FIG. \ref{fig:topo_bands}a(ii)). The band gap closing signals a topological phase transition, as it is at this point that the topological invariant can change value. In this case, $\mathcal{H}$ and $\mathcal{H}'$ exist in different topological phases. If the gap remains open during the transformation then $\mathcal{H}$ and $\mathcal{H}'$ remain in the same topological phase. The gap remaining open is often enforced by a system symmetry, and so in order for a topological phase transition to occur a symmetry breaking must occur. If the symmetry is preserved then no topological phase transition will take place. 

The topological invariant described above is a bulk quantity (and is intimately linked to the bulk Hamiltonian). For an insulating material with a non-zero topological invariant surrounded by vacuum, we have a boundary on the interface of these two insulators with differing topological order parameters. {\it The change in topological invariant at the boundary requires the band gap to close, whilst remaining gapped in the bulks of both media.} This results in localized boundary states, which necessarily traverse the band gap. Physically, the picture we then have is of an insulating bulk, with conducting states localized on the boundary of the material as illustrated in FIG. \ref{fig:topo_bands}b. Many materials with topological band structures (such as HgTe/CdTe wells~\cite{konig2007quantum} and Bi$_2$Se$_3$ family materials~\cite{hsieh2009observation,chen2009experimental,zhang2009topological}) owe their topological properties to spin-orbit coupling, causing an inverted band order and subsequently for edge states to appear in the gap (see FIG. \ref{fig:topo_bands}c).

\subsection{Time-reversal symmetry and TIs}  
\label{subsec:TIs}

The pioneering example of topology in condensed matter is the 2D quantum Hall (QH) effect~\cite{klitzing1980new,thouless1982quantized} (schematically illustrated in FIG. \ref{fig:topo_history}). For a 2D electronic system at low temperature subjected to large magnetic fields, the Hall resistivity $\rho_{xy}$ exhibits plateaus, characterized by a topological invariant known as the Chern number, which takes integer values. The large magnetic field results in electrons in the bulk being localised in small cyclotron orbits, whilst electrons at the edge experience truncated cyclotron orbits and travel along the edge of the system giving conducting edge states. The time-reversal symmetry in this system is explicitly broken by the magnetic field. 

Conversely to the case of the QH state, in QSH and TIs, conducting edge states are only present when TR symmetry is preserved, such that the fermionic time-reversal operator $T_f$ commutes with the system Hamiltonian $[\mathrm{\mathcal{H}}, T_f] = 0$, and the fermion condition  $T_f^2 = -1$, is obeyed. The TR symmetry enforces Kramers degeneracy, that is to say that for every eigenstate $|n\rangle$, its time-reversed partner $T_f | n \rangle$ is also an eigenstate and has the same energy, but is orthogonal. This can be demonstrated using the anti-unitary nature of $T_f$ and the fermion condition, such that 
\begin{equation}
\begin{split}
\langle n , T_f  n \rangle &= \langle T_f n , T_f^2  n \rangle^* =  -\langle T_f n , n \rangle^* = -\langle  n , T_f n \rangle, 
\end{split}
\end{equation}
and so $\langle  n , T_f n \rangle = 0$. {\it Consequently, in the presence of time-reversal symmetry these counter-propagating states cannot backscatter into one another.} If a TR breaking perturbation is applied (such as a magnetic impurity in a topological insulator), a gap will open within the dispersion relation of the edge states and they will no longer conduct, reducing the TI to a trivial insulator. However if a non-TR breaking perturbation is applied the states will remain gapless and will be robust against the perturbation. 

\section{Topological photonics} 
\label{sec:topo_photonics}

When discussing topological photonics, there are two distinct topics to consider. The first is that of topological electronic systems and their interaction with photons~\cite{siroki2016single,PhysRevB.97.115420,di2013observation,Stauber2017}, and the second is that of purely photonic systems with a non-trivial topology~\cite{ozawa2018topological,lu2014topological,khanikaev2017two,Sun2017}. In this second area, we have photonic analogue systems that aim to mimic the band structure and topological properties of a known electronic topological system, as well as topological photonic systems with no electronic counterpart.

Photonic platforms that can support topological systems exist over a vast range of frequencies, however miniaturization to the nanoscale still proves difficult as some of the platforms are fundamentally limited in size, or if there are no fundamental issues then there are technical hurdles to overcome in downsizing the systems. We review some major milestones in topological photonics and the challenges of their corresponding platforms. 

\begin{figure}[t]
\includegraphics[width=\columnwidth]{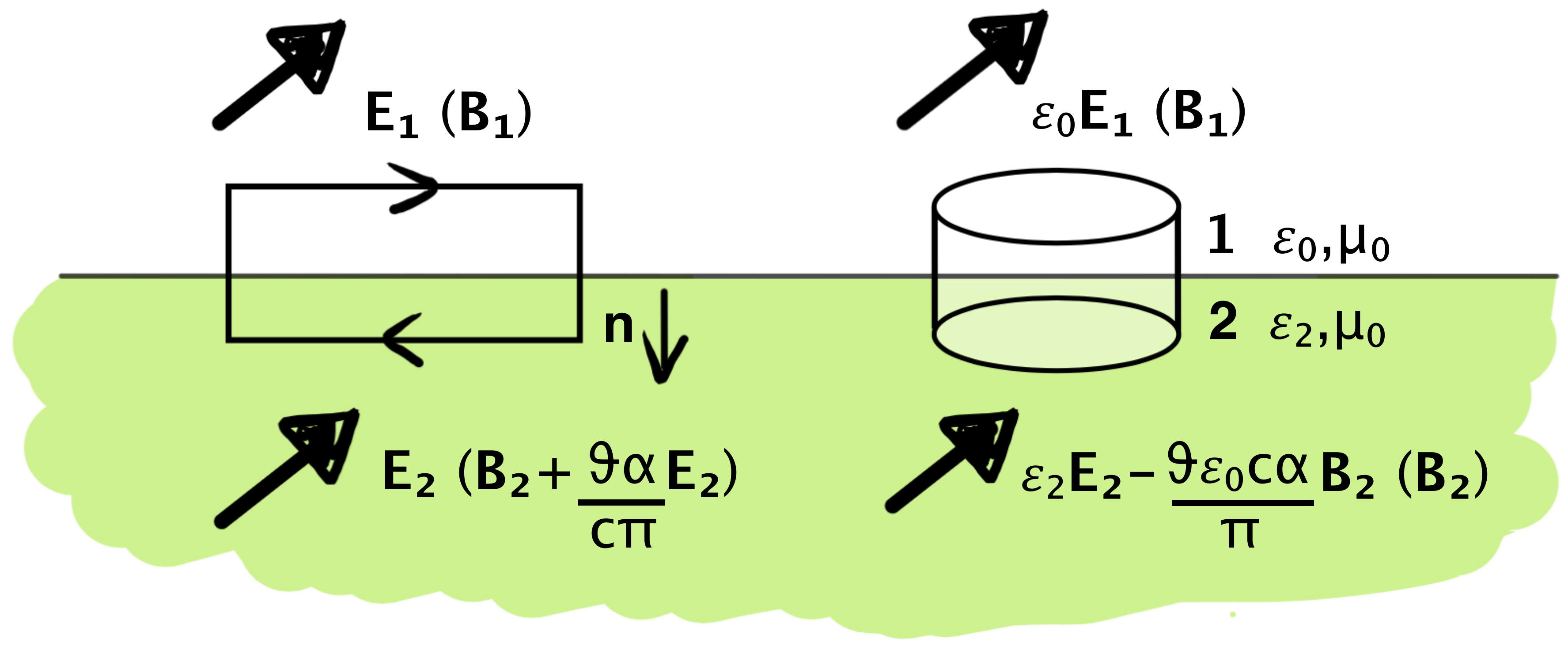}
\caption{\textbf{Topologically modified Maxwell's equations} Boundary conditions at the surface of a topological insulator.\label{fig:mod_topo}}
\end{figure}

\subsection{Topological insulators interacting with light} \label{subsec:tis_with_light}

We review 3D topological insulators (as introduced in section \ref{subsec:TIs}) and their response when irradiated with light. Due to the presence of TR symmetry, a new term can be added to the action of the system ~\cite{RevModPhys.83.1057}, given by 
\begin{equation}
S({\vartheta}) = -\frac{\alpha}{16 \pi \mu_0} \int d^3x dt \vartheta(x,t)\epsilon^{\mu \nu \rho \tau} F_{\mu \nu} F_{\rho \tau},
\end{equation}

where $\mu_0$ is the permeability of free space, $\alpha$ is the fine structure constant and $\epsilon^{\mu \nu \rho \tau}$ is the fully anti-symmetric 4D Levi-Civita tensor. $F_{\mu \nu}$ is the electromagnetic tensor and $\vartheta(x,t)$ is an angular variable which we assume to be constant in order to preserve spatial and temporal translation symmetry. Normally this term would not be considered in a theory of electromagnetism as it does not conserve parity, which the electromagnetic interaction is known to do. However if $\vartheta$ is defined modulo $2\pi$ and is restricted to take the values $0$ and $\pi$ only, we find a theory that conserves both parity and time reversal symmetry.  These two values of $\vartheta$ give us either a topologically-trivial insulator ($\vartheta=0$) or a topological insulator ($\vartheta=\pi$), where $\vartheta$ is the  topological invariant. Expanding, this leads to a new term in the Lagrangian $\propto \mathbf{E} \cdot \mathbf{B}$. This additional term in the Lagrangian leads to modified Maxwell equations 
\begin{equation}
\begin{split}
\mathbf{\nabla}\cdot \mathbf{D} &= \rho +  \frac{\alpha \epsilon_0 c}{\pi}  \left( \mathbf{\nabla} \vartheta \cdot \mathbf{B} \right) ,
\\ \mathbf{\nabla}\times \mathbf{H} - \frac{\partial \mathbf{D}}{\partial t} &= \mathbf{j} -\frac{ \alpha }{\pi \mu_0 c}  \left( \mathbf{\nabla} \vartheta \times \mathbf{E} \right) , 
\\ \mathbf{\nabla} \times \mathbf{E} + \frac{\partial \mathbf{B}}{\partial t} &= 0,
\\ \mathbf{\nabla} \cdot \mathbf{B} &= 0,
\end{split}
\end{equation}
where $\mathbf{D} = \epsilon \mathbf{E}$ and $\mathbf{H} = \frac{1}{\mu}\mathbf{B}$. To write $\mathbf{D}$ and $\mathbf{H}$ in these forms we have assumed a linear material such that polarisation is given by $\mathbf{P}=\epsilon_0 \chi \mathbf{E}$ and magnetization by $\mathbf{M} = \chi_m \mathbf{H}$ where $\chi$ and $\chi_m$ are electric and magnetic susceptibility respectively. $\mathbf{H}$ is the magnetic field and $\mathbf{D}$ is electric displacement. $\epsilon$ is the material permittivity, $\mu$ is the material permeability (which for all materials we will cover in this work is given by $\mu = \mu_0$). $\mathbf{j}$ describes free currents, whilst $\rho$ gives the charges. To be expected, when in a topologically trivial phase (such that $\vartheta = 0$) the equations reduce to the ordinary Maxwell equations.  As an alternative to the topologically modified Maxwell equations, we can also write the usual Maxwell equations with modified constituent equations~\cite{RevModPhys.83.1057}, such that 
\begin{equation}
\begin{split}
\mathbf{D} & = \varepsilon \mathbf{E} - \frac{\vartheta \varepsilon_0 c \alpha}{\pi}  \mathbf{B},
\\ \mathbf{H} & = \frac{1}{\mu_0}\mathbf{B} + \frac{\vartheta \alpha}{\pi \mu_0 c}\mathbf{E}.
\end{split}
\end{equation}
There have been various successful theory and experimental proposals on this topic~\cite{maciejko2010topological,tse2010giant,0953-8984-26-12-123201,di2013observation}, and in box~1 we include a new proposal in which a layer of topological insulator material is added to an Otto surface-plasmon-resonance (SPR) configuration in order to show that the conducting surface state of the TI can support a plasmon mode. This is due to the unique $\mathbf{E}\cdot\mathbf{B}$ coupling of the system and the correspondent new boundary conditions (see FIG.\ref{fig:mod_topo}). A plasmon mode excited with purely {\it p}-polarised incoming light will rotate out of the plane of the material and a small component of {\it s}-polarised light will be transmitted or reflected (see box~1). We direct readers to the prospective paper on experimental methods for creating films of 3D TI materials and the existence of 2D Dirac plasmons in these systems ~\cite{ginley2018dirac}. 

The elegant method of using modified Maxwell equations to characterize the optical properties of TIs is valid when the surface states are well described classically (i.e. long wavelength), which holds well for a TI slab as in the study in box 3. In section \ref{sec:topo_nano} we discuss spherical TI particles. For nanoparticles (with particle radius $R<100$nm) a classical description of the surface states is no longer valid, and we must invoke quantum mechanics to fully describe them and their interaction with incoming light~\cite{siroki2016single}. 


\begin{tcolorbox}[float*,colbacktitle=teal!50!white, coltitle=black,colback=teal!10!white, width=\textwidth, title=\textbf{BOX 1: Exploring Dirac Plasmons with Prism Coupling}]

It is instructive to see how an extremely widespread system in nanophotonics can be used to study TIs and in particular the excitation of Dirac plasmons. A typical prism coupling system is shown in FIG. B1a. A prism (Silicon, $n=3.5$) is positioned on top of a dielectric spacer (e.g. Benzocyclobutene~\cite{PERRET20082276} with $\epsilon=2.5$ and thickness $200\mu$m), a thin film of a topological insulator (Bi$_2$Se$_3$ with $\epsilon=-3.4+36.7i$ at 1 THz~\cite{siroki2016single}, thickness 8$\mu$m) all on top of a substrate with the same refractive index of the spacer. The light coming from the prism at a grazing angle will excite surface modes in the film below the spacer by a tunneling effect. Such a system is widely used in plasmonics, where a metal film replaces the TI~\cite{giannini2011plasmonic}. In such experiments, abrupt reduction of the reflectivity is observed for angles of incidence in the zone of the total reflection. In fact, with the right conditions it is possible to convert the entirety of the incident light into a surface plasmon polariton (SPP)~\cite{rivas2008surface}.
\begin{wrapfigure}[19]{r}{0.6\textwidth}
\begin{tcolorbox}[width=0.6\textwidth, colback=white]
    \includegraphics[width=\textwidth]{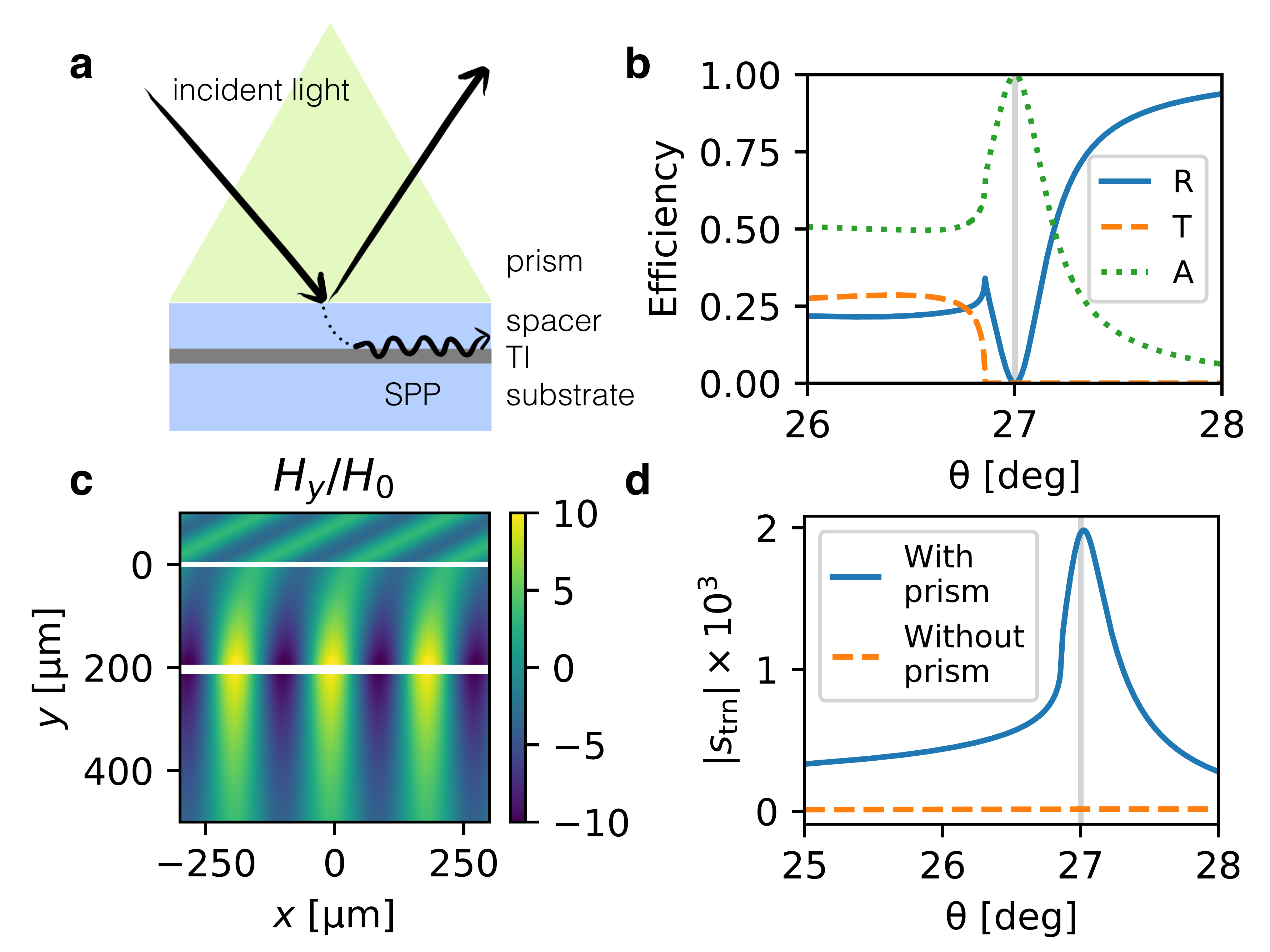}    
    FIG B1: \textbf{Topological SPP} \textbf{(a)} Otto configuration comprised of a prism, spacer and topological insulator film on a substrate. \textbf{b} Reflection, transmission and absorption. \textbf{(c)} H near-field for {\it p}-polarised light.  \textbf{d} Transmitted {\it s}-polarised light with and without prism.
\end{tcolorbox}
\end{wrapfigure} 
\\ Here, we utilise the same idea but instead of a SSP, we excite something similar to a surface exciton-polariton in an absorbing media, i.e. the same phenomenon as observed for Silicon in the near UV~\cite{giannini2008long}. We utilise the high absorption due to the $\alpha$-phonon in the Bi$_2$Se$_3$~\cite{siroki2016single}.
The calculation can be done easily by modifying the boundary conditions to include topology in the multilayer system~\cite{note1}. Following from the theory in section \ref{subsec:tis_with_light} we enforce the following topological boundary conditions~\cite{RevModPhys.83.1057}, which are given diagrammatically in FIG. \ref{fig:mod_topo}, 

\begin{equation}
\begin{split} 
\mathbf{E}_2^{\perp} &= \frac{\mathbf{E}_1^{\perp}}{\varepsilon_2^r} + \frac{\vartheta c \alpha}{\varepsilon_2^r \pi} \mathbf{B}_2^{\perp}
\\ \mathbf{E}_2^{\parallel} &=\mathbf{E}_1^{\parallel}
\\ \mathbf{B}_2^{\perp} &= \mathbf{B}_1^{\perp} 
\\ \mathbf{B}_2^{\parallel} &= \mathbf{B}_1^{\parallel} - \frac{\vartheta \alpha}{c \pi} \mathbf{E}_2^{\parallel}.
\end{split}
\end{equation}

We remark that the above boundary conditions mix the polarisations. For example, a p-polarized incident light can be partially converted in s-polarization after the interaction with the TI. In other words, the TI introduces a magneto-optical effect and a rotation of the light polarization. The observation of the magneto-optical effect using the prism coupling method could offer a new way to study topological insulators. The result can be seen in FIG. B1b where we show reflectivity (blue), transmission (orange dashed) and absorption (green dotted) for a p-polarized incident field at 1 THz. We can see that a sharp peak in the absorption is obtained at $\sim$27 degrees. At this angle, in the total reflection zone, we excite a surface polariton in the thin film, as can also be seen from the near fields in FIG. 1Bc. 
Let us now analyze the light converted from {\it p}-polarisation to {\it s}-polarisation. We see a strong conversion at the surface polariton angle of excitation (blue curve in FIG. B1d). In order to analyze the enhancement of such a conversion, we also plot the light converted to {\it s}-polarisation when the prism is not present, i.e. when we do not excite the surface polariton.
\end{tcolorbox}

\subsection{Topological photonic analogues} \label{subsec:topo_photonic_analogues}

So far we have discussed merging the physics of topological electronic systems with light, but we now move to the concept of topological photonic analogues, in which we aim to mimic the properties of topological electronic structures using bosonic degrees of freedom (i.e. photons). We begin by discussing how to construct photonic band structures, before moving onto specific examples both with and without time-reversal symmetry. 

\begin{table*}
    \centering
    \begin{tabular}{l l l}
    \hline
     & $\quad\quad$Quantum Mechanics & Electrodynamics  \\
     \hline
     \noalign{\vskip 2mm}
      Field   &  $\quad\quad\Psi (\mathbf{r},t) = \Psi (\mathbf{r}) e^{-i\frac{Et}{\hbar}}\quad\quad$  & $\mathbf{H} (\mathbf{r},t) = \mathbf{H} (\mathbf{r}) e^{-i\omega t}$  \\
  Eigenvalue problem   & $\quad\quad\mathcal{H} \Psi = E \Psi$ & $\Theta \mathbf{H} = \left(\frac{\omega}{c}\right)^2 \mathbf{H}$  \\
     Hermitian operator    & $\quad\quad\mathcal{H} = -\frac{\hbar^2}{2m}\nabla^2 + V(\mathbf{r})$ & $\Theta = \nabla \times \frac{1}{\varepsilon(\mathbf{r})} \nabla \times $ \\ 
     \noalign{\vskip 2mm}
     \hline
    \end{tabular}
    \caption{\textbf{QM vs EM} Quantities of quantum mechanics alongside their photonic analogue counterparts.}
    \label{tab:QM_vs_EM}
\end{table*}

\subsubsection{Photonic band structures}
\label{subsubsec:photonic_bands}

The first hurdle in emulating topological electronic structures with photonic systems is by devising a way to create a band structure for light. Electrons in vacuum have a gapless, parabolic dispersion relation, but when presented with a periodic potential such as a crystal lattice, gaps may open in which the electrons will not-propagate, i.e. we have a photonic crystal~\cite{Yablonovitch1987,John1987}. Similarly, photons in vacuum exhibit a gapless, linear dispersion relation which can become gapped on the introduction of a periodic medium. We overview this analogy between the Maxwell equations in a periodic medium and quantum mechanics with a periodic Hamiltonian~\cite{joannopoulos2011photonic}.

We begin with the macroscopic Maxwell equations, 
\begin{equation}
    \begin{split}
    \nabla \cdot \mathbf{H}(\mathbf{r},t) = 0 \quad \quad \nabla \times \mathbf{H}(\mathbf{r},t) - \varepsilon_0 \varepsilon(\mathbf{r}) \frac{\partial \mathbf{E}(\mathbf{r},t)}{\partial t} = 0
    \\ \nabla \cdot \left[ \varepsilon(\mathbf{r})\mathbf{E}(\mathbf{r},t) \right] = 0 \quad \quad \nabla \times
\mathbf{E}(\mathbf{r},t) + \mu_0 \frac{\partial \mathbf{H}(\mathbf{r},t)}{\partial t} = 0. 
    \end{split}
\end{equation}
written in terms of $\mathbf{H}$ and $\mathbf{E}$ fields, where both fields are dependent on both $\mathbf{r}$ and $t$.  We restrict ourselves to real and positive $\epsilon (\mathbf{r})$ and linear materials (although the theory can be generalized for the presence of loss). For mathematical convenience we write the fields as complex-valued fields, such that we have a spatial-mode profile multiplied by a time-dependent complex exponential, 
\begin{align}
    \mathbf{H}(\mathbf{r},t) &= \mathbf{H} (\mathbf{r}) e^{-i \omega t} 
    \\ \mathbf{E}(\mathbf{r},t) &= \mathbf{E} (\mathbf{r}) e^{-i \omega t} ,
\end{align}
with the proviso that we must take only the real part of the fields when we wish to recover the physical fields. In doing so, we can substitute into the Maxwell equations and combine into a single eigenvector equation for $\mathrm{\mathbf{H}}(\mathbf{r})$,
\begin{equation}
    \nabla \times \left( \frac{1}{\varepsilon (\mathbf{r})} \nabla \times \mathbf{H} (\mathbf{r})\right) = \left( \frac{\omega}{c}\right)^2 \mathbf{H} (\mathbf{r}),
\end{equation}
in which we can write the Hermitian operator $\Theta$ as everything that acts on $\mathrm{\mathbf{H}}(\mathbf{r})$ on the left hand side of the equation, such that 
\begin{equation} \label{eq:eigen}
    \Theta \mathbf{H} (\mathbf{r}) \equiv  \nabla \times \left( \frac{1}{\varepsilon (\mathbf{r})} \nabla \times \mathbf{H} (\mathbf{r})\right).
\end{equation}
The field, eigenvector equation and Hermitian operator are repeated in table \ref{tab:QM_vs_EM}, in which they are compared to their counterparts in quantum mechanics~\cite{joannopoulos2011photonic}. 

In order to create a band structure from this eigenvalue problem, we now introduce a discrete translation symmetry of the material (equivalent to demanding periodic boundary conditions). Mathematically, this results in $\varepsilon (\mathbf{r}) = \varepsilon (\mathbf{r}+\mathbf{R})$, where $\mathbf{R}$ is an integer multiple of the lattice step vector (i.e. the vector traversed before the system pattern repeats). The field $\mathrm{\mathbf{H}}(\mathbf{r})$ can still be considered as a plane wave, but now modulated by a periodic function due to the periodicity of the lattice, such that 
\begin{equation} \label{eq:bloch}
    \mathbf{H}_\mathbf{\mathbf{k}} (\mathbf{r}) = e^{i \mathbf{k} \cdot \mathbf{r}} \mathbf{u}_{\mathbf{k}} (\mathbf{r}), 
\end{equation}
where $\mathbf{u}(\mathbf{r})$ is a function with the same periodicity as the lattice and $\mathbf{k}$ is the Bloch wave vector, a conserved quantity so long as the discrete translation invariance of the system holds. This result is known as Bloch's theorem and is analogous to the periodicity of the electron wave function in a crystal lattice.  Putting equations \ref{eq:eigen} and \ref{eq:bloch} together, we obtain a new eigenproblem for $\mathbf{u}_{k} (\mathbf{r})$,

\begin{equation}
    \Theta_\mathbf{k} \mathbf{u}_{\mathbf{k}} (\mathbf{r}) = \left(\frac{\omega(\mathbf{k})}{c}\right)^2 \mathbf{u}_{\mathbf{k}} (\mathbf{r}),
\end{equation}
where the new eigenoperator is given as 
\begin{equation}
    \Theta_{\mathbf{k}} \equiv (i \mathbf{k}+\nabla) \times \frac{1}{\varepsilon (\mathbf{r})} \left( i \mathbf{k} +\nabla \right) \times.
\end{equation}

\begin{figure*}
\includegraphics[width=\textwidth]{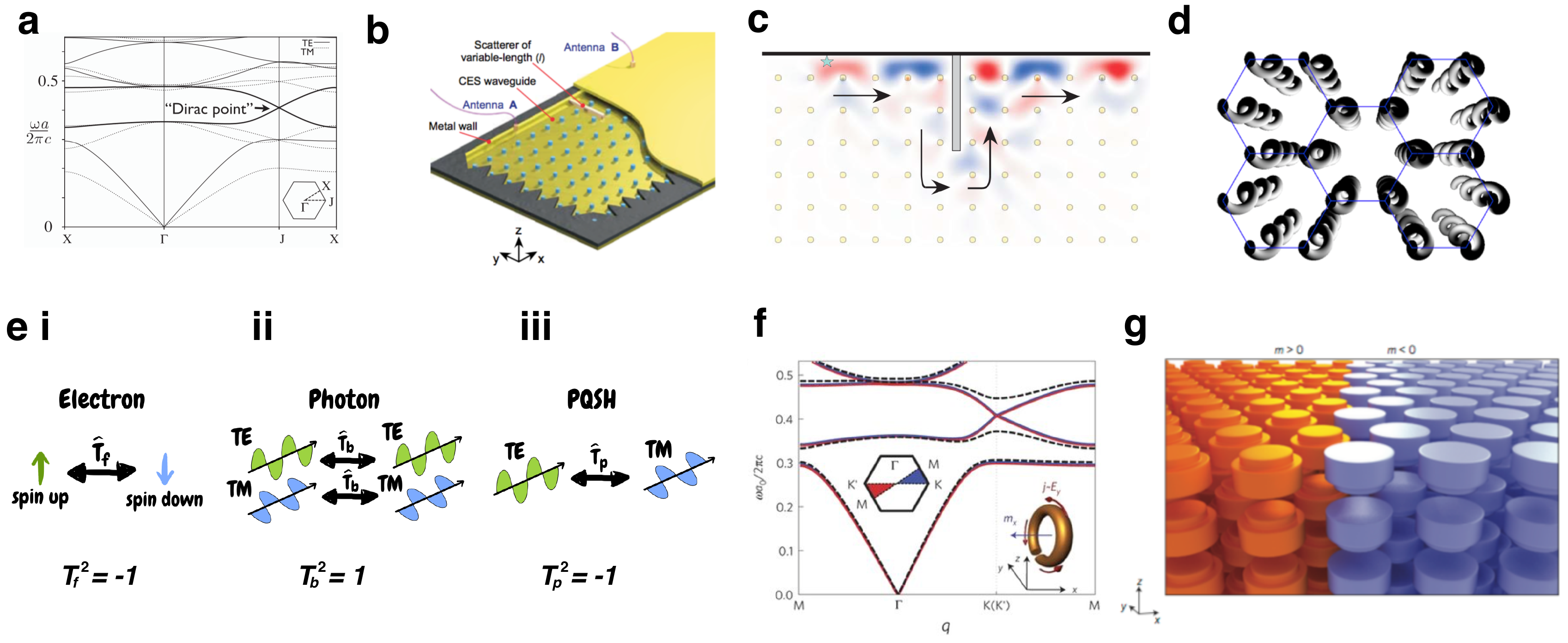}
\caption{\textbf{Proposals and realisations of topological photonic systems} \textbf{(a)} The first proposal of PQH state, reprinted figure with permission from~\cite{haldane2008possible} Copyright (2008) by the American Physical Society. \textbf{(b)} The first experimental realisation of PQH and \textbf{(c)} the edge states from this work, both figures reprinted by permission from~\cite{wang2009observation}, Springer Nature. \textbf{(d)}  Proposal and experimental demonstration of a photonic Floquet topological insulator, reprinted with permission from ~\cite{rechtsman2013photonic}, Springer Nature. \textbf{(e)} Time-reversal symmetry for (i) electrons, (ii) photons, and (iii) a scheme for pseudo-fermionic TR symmetry. \textbf{(f)} First proposal for a 3D photonic TI using the pesudo-fermionic time reversal scheme shown in (e), reprinted with permission from ~\cite{khanikaev2013photonic}, Springer Nature. \textbf{(g)} The first proposal for 3D all-dielectric photonic TI, reprinted with permission from ~\cite{slobozhanyuk2017three}, Springer Nature. \label{fig:lit_review}}
\end{figure*}

Media with a periodic dielectric function can be manufactured in a multitude of ways. Photonic crystals may be formed of periodic dielectric, metallo-dielectric, gyroelectric or gyromagnetic structures. To diffract electromagnetic waves of a given wavelength $\lambda$, the periodicity of the photonic crystal structure must be $\approx \frac{\lambda}{2}$. In order to diffract visible light ($400\mathrm{nm}<\lambda<700$nm) the photonic nanostructure becomes increasingly more difficult to construct. Photonic devices made of arrays of optical resonators and coupled waveguides are also of interest, but again miniaturization is a challenging goal.  

When incorporating topology into the photonic band structure, we mirror the concepts found in topological condensed matter systems. We focus on the two main types of system: those in which time-reversal symmetry is explicitly broken in order to support topological states and those in which we aim to preserve time-reversal symmetry.

\begin{tcolorbox}[float*,colbacktitle=teal!50!white, coltitle=black,colback=teal!10!white, width=\textwidth, title=\textbf{BOX 2: Chiral symmetry and the SSH model}]
Here we discuss the Su-Schrieffer-Heeger (SSH) model ~\cite{PhysRevLett.42.1698,asboth2016short}, whose topological properties are linked to sublattice symmetry. We study a 1D chain of atoms with nearest-neighbour interactions only, in which bond strength between atoms alternates (illustrated in FIG. B2a). This is a single electron Hamiltonian, written as  
\begin{equation}\begin{split}\mathcal{H} = v \sum_{m=1}^N  |m,B\rangle \langle m,A| +  w \sum_{m=1}^{N-1} |m+1,A\rangle \langle m,B| + h.c.  ,\end{split}\label{eq:SSH_ham}\end{equation} 
where for $m \in \{1,2,...,N\}$, $|m,A\rangle$ and $|m,B\rangle$ are the states for which the electron is on unit cell $m$, on either sublattice $A$ or $B$ respectively. We define projection operators for each sublattice $P_A$, $P_B$ and the sublattice operator $\Sigma_z$, 
\begin{equation}
P_A = \sum_{m=1}^N |m,A \rangle \langle m,A |, \quad P_B = \sum_{m=1}^N |m,B \rangle \langle m,B |, \quad \mathrm{and} \quad  \Sigma_z = P_A - P_B.
\end{equation}

\begin{wrapfigure}{r}{0.4\textwidth}
\begin{tcolorbox}[width=0.4\textwidth, colback=white]
\begin{center}
      \includegraphics[width=\textwidth]{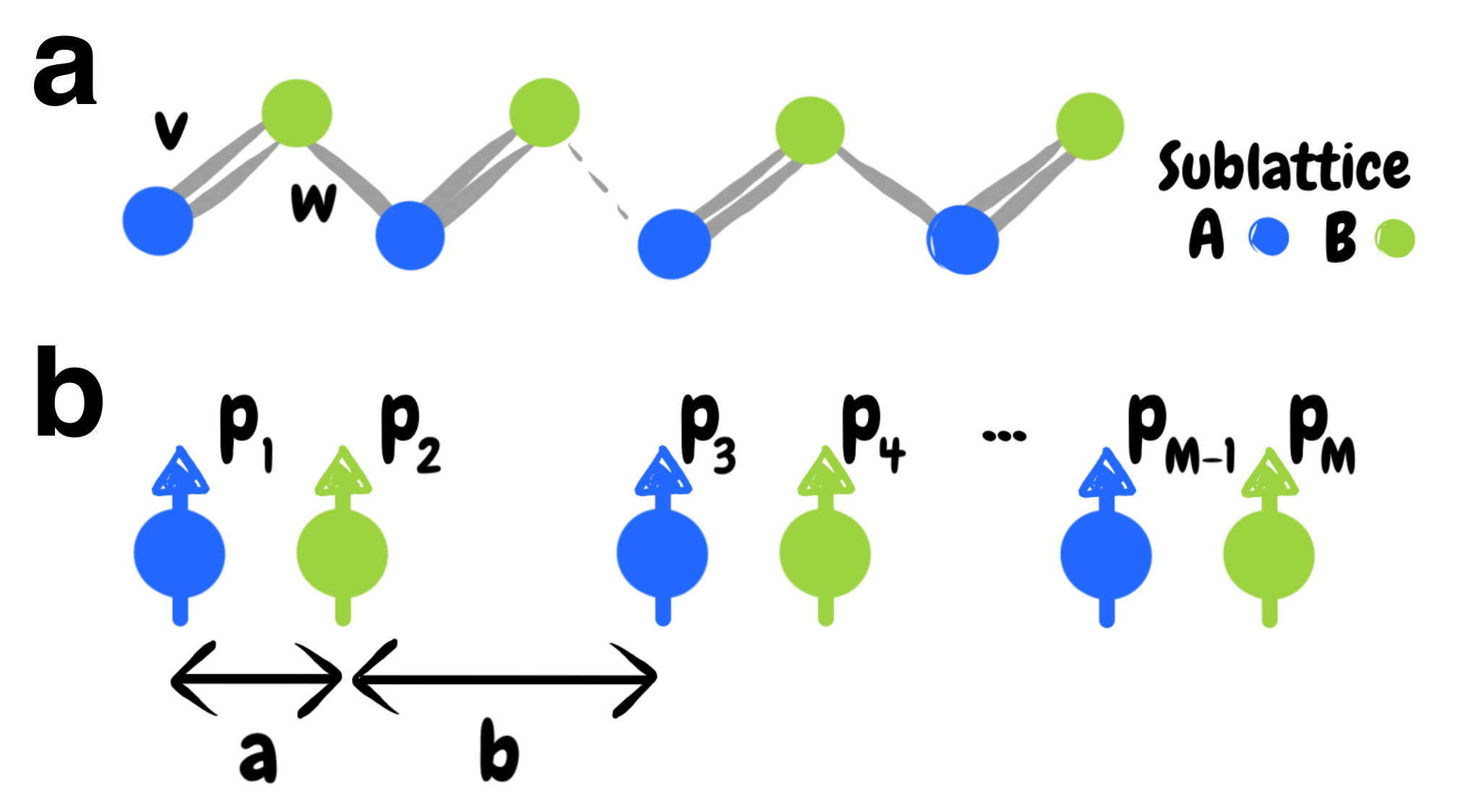}      
\end{center}
    FIG B2: \textbf{Schematic of SSH model} \textbf{(a)} The SSH model \textbf{(b)} Dipole analogue of SSH model
\end{tcolorbox}
\end{wrapfigure}

Applying the sublattice operator to the Hamiltonian we see that $\Sigma_z \mathcal{H} \Sigma_z = - \mathcal{H}$. This relationship holds for inversion of the sublattices such that $P_A \rightarrow P_B$ and $P_B \rightarrow P_A$. This tells us that if $|\psi_A,\psi_B\rangle$ is an eigenstate of the system with energy $E$,  $|\psi_A,-\psi_B\rangle$ will be an eigenstate with energy $-E$.  This means that for a system with band gap around 0, there will be an equal number of states below the gap as above the gap and this will can only be violated if the gap closes, causing a topological phase transition. 

A photonic analogue of the SSH model can quite simply be envisioned by studying a chain of dipoles~\cite{Ling:15} with dipole moments $\mathbf{p}_n$ and which are alternately spaced by distances $a$ and $b$ (as shown in FIG. B2b), described by the coupled dipole equations
\begin{equation}
    \frac{1}{\alpha (\omega)} \mathbf{p}_i = \sum^M_{i \neq j} G (\mathbf{r}_{ij}, \omega) \mathbf{p}_j,
\end{equation}
where $G(\mathbf{r}_{ij},\omega)$ is the matrix-valued Green's function, depending on the separation of the dipoles $\mathbf{r}_{ij} = \mathbf{r}_i-\mathbf{r}_j$ and $\alpha (\omega)$ is the polarisability of whichever platform is being used to form the dipoles. In the electrostatic, nearest-neighbour limit this system will obey the SSH Hamiltonian. 
\end{tcolorbox}

\subsubsection{Explicit time-reversal breaking}
\label{subsubsec:explicit_TR_breaking}

The first theoretical proposal for a photonic analogue of the quantum Hall effect came in 2008 ~\cite{haldane2008possible,PhysRevA.78.033834}, in which the interface between gyroelectric photonic crystals of differing Chern number was studied. The proposed system comprised of a hexagonal array of dielectric rods exhibiting a Faraday effect, with the Faraday-effect enabling the time-reversal breaking and opening the band gap. The interface between the photonic crystals creates a domain wall across which the direction of the Faraday axis reverses. The Faraday effect vanishing at the domain wall results in Dirac-like edge states at this point (as illustrated in FIG.~\ref{fig:lit_review}a). These unidirectional photonic modes are the direct analogue of chiral edge states in a quantum Hall system. While this work was limited to photonic band structures containing Dirac points, another study ~\cite{PhysRevLett.100.013905} noted that a Dirac cone is not imperative for a system to support edge states, merely that the band structures of materials of either side of an interface have different Chern numbers of bands below the gap. They put forward a new proposal using a square lattice gyromagnetic crystal operating at a microwave frequency, resulting in time-reversal breaking strong enough for the effect to be easily measured. Both proposals also note that the phenomenon of chiral edge states should be independent of the underlying particle statistics as the Chern number is defined in terms of single-particle Bloch functions. The number of chiral edge states is equal to the sum of the Chern numbers of all bands below the band gap, and the Chern number may only be non-zero if the system explicitly breaks time-reversal symmetry. 

The first experimental observation of photonic chiral edge states came in 2009 ~\cite{wang2009observation}, utilising a magneto-optical photonic crystal in the microwave regime. The gyromagnetic, 2D-periodic photonic crystal consisted of a square lattice of ferrite rods in air bounded on one side by a non-magnetic metallic cladding to prohibit radiation leakage (illustrated in FIG. \ref{fig:lit_review}b). The experimental study demonstrated unidirectional edge states (shown in FIG. \ref{fig:lit_review}c)which were robust against scattering from disorder, even in the presence of large metallic scatterers. An experimental realisation using a 2D honeycomb array of ferrite rods in 2011 ~\cite{poo2011experimental} showed that an auxillary cladding is not necessary as edge states can be constructed such that they necessarily evanesce in air. The gyromagnetic effect employed in this system is limited by the Larmor frequency of the underlying ferrimagnetic resonance, which is on the order of tens of gigahertz. 

Using a different approach, Floquet topological insulators induce topological effects not by explicitly breaking time-reversal symmetry, but by time-periodic modulation (driving). Topologically protected edge states may arise at the boundary of two Floquet systems much in the same way that they would in the usual topological insulators. Photonic Floquet topological insulators were first realized in 2013 ~\cite{rechtsman2013photonic}, using a platform of coupled helical waveguides (whose cross section have diameters $\sim\mu$m, arranged in a honeycomb lattice structure (as shown in FIG. \ref{fig:lit_review}d). The helicity of the waveguides break inversion symmetry in the z direction (the axis of propagation), which acts analogously to the breaking of TR symmetry in a solid state system, which is evident from the equivalence between the paraxial wave equation in EM and the Schr\"{o}dinger equation in QM. The structure results in topologically protected, unidirectional edge states. 

\subsubsection{Fermionic pseudo-time-reversal symmetry and photonic TIs}
\label{subsubsec:photonic_TIs}

The photonic analogues of the quantum Hall effect outlined above exhibit photonic edge states which are topologically protected, but can be challenging to manifest experimentally as a strong magnetic field is usually needed. Systems in which time-reversal symmetry is not broken do not require external biasing such as an applied magnetic field, but can pose their own challenges. We now discuss photonic analogues of quantum spin Hall states and topological insulators. As covered in section \ref{subsec:TIs}, the time-reversal invariance in electronic systems is intimately linked to the fermionic condition $T_f^2=-1$, and the topological protection from backscattering directly emerges from this condition. Photonic systems are constructed from bosonic degrees of freedom (namely photons) and bosonic time-reversal symmetry (which obeys the condition $T_b^2 = 1$) does not give protection from back-scattering. In photonic systems has been proposed instead, to construct a pseudo-fermionic time-reversal operator $T_P$, by combining the bosonic time-reversal operator with some other symmetry of the structure, for example a crystal symmetry~\cite{PhysRevLett.114.223901}. The platforms we will now discuss construct $T_P$ either with the use of polarisation degeneracy or by relying on a lattice symmetry. 

The first theoretical proposal for a photonic QSH state came in 2011~\cite{hafezi2011robust}, using a two-dimensional array of coupled resonator optical waveguides (CROW). Degenerate clockwise and anti-clockwise modes of a 2D magnetic Hamiltonian behave analogously to spins with spin-orbit coupling in the electronic quantum spin Hall effect.  The proposal was then realised experimentally in 2013~\cite{hafezi2013imaging}, and another proposal using optical ring-resonators followed soon after, eradicating the need to fine-tune the inter-resonator couplings~\cite{liang2013optical}.

The first metamaterial proposal for a photonic analogue of a Z$_2$ topological insulator was presented in 2013~\cite{khanikaev2013photonic}, using a metacrystal formed of a 2D superlattice of subwavelength metamaterials. The spin degeneracy leading to pseudo-fermionic time reversal invariance is constructed by enforcing $\epsilon = \mu$, resulting in TE and TM modes in the system propagating with equal wavenumbers. This allows one to write linear combinations of the fields which propagate with equal wavenumber and are double degenerate. These states are connected with a pseudo-fermionic time reversal operator (as depicted in FIG. \ref{fig:lit_review}e) and will act analogously to spin degenerate states in an electronic system. The photonic band structure of such a system is shown in FIG. \ref{fig:lit_review}f. 

Photonic topological crystalline insulators obey pseudo-fermionic TR symmetry enforced by the bosonic TR symmetry of the photons and a crystal symmetry. In 2015 a purely dielectric scheme was proposed~\cite{wu2015scheme}, which does not require a magnetic field and is constructed from cylinders in a honeycomb lattice which is distorted such that a triangular lattice emerges with a hexagon of cylinders at each site. The system has $C_6$ symmetries and helical edge states. This structure was used in 2017 to experimentally demonstrate that deep subwavelength scale topological properties can be produced in crystalline metamaterials due to multiple resonant scattering~\cite{yves2017crystalline}.  

It is interesting to note, that finite crystals using the same scheme of dielectric cylinders form photonic topological insulator particles present modes that can are topological whispering gallery modes, as was first theoretically proposed~\cite{siroki2017topological} and later experimentally realized~\cite{Yang2018}. Rather than a continuous spectrum of edge states as seen in an infinite system, these particles exhibit a discrete spectrum of edge states which (like the infinite system) support unidirectional, pseudospin-dependent propagation. The discrete nature of the edge states agrees with the observation of discrete peaks in experimental transmission measurements. Actually, in any experimental realization of a topological photonic crystal we have a low number of unit cells compared with the Avogadro number of electrons in a electronic topological insulator; this has an important repercussion in the number of discrete states, i.e. in any realistic experiment with a topological photonic crystal only few states are possible.

\begin{figure}[t]
\includegraphics[width=\columnwidth]{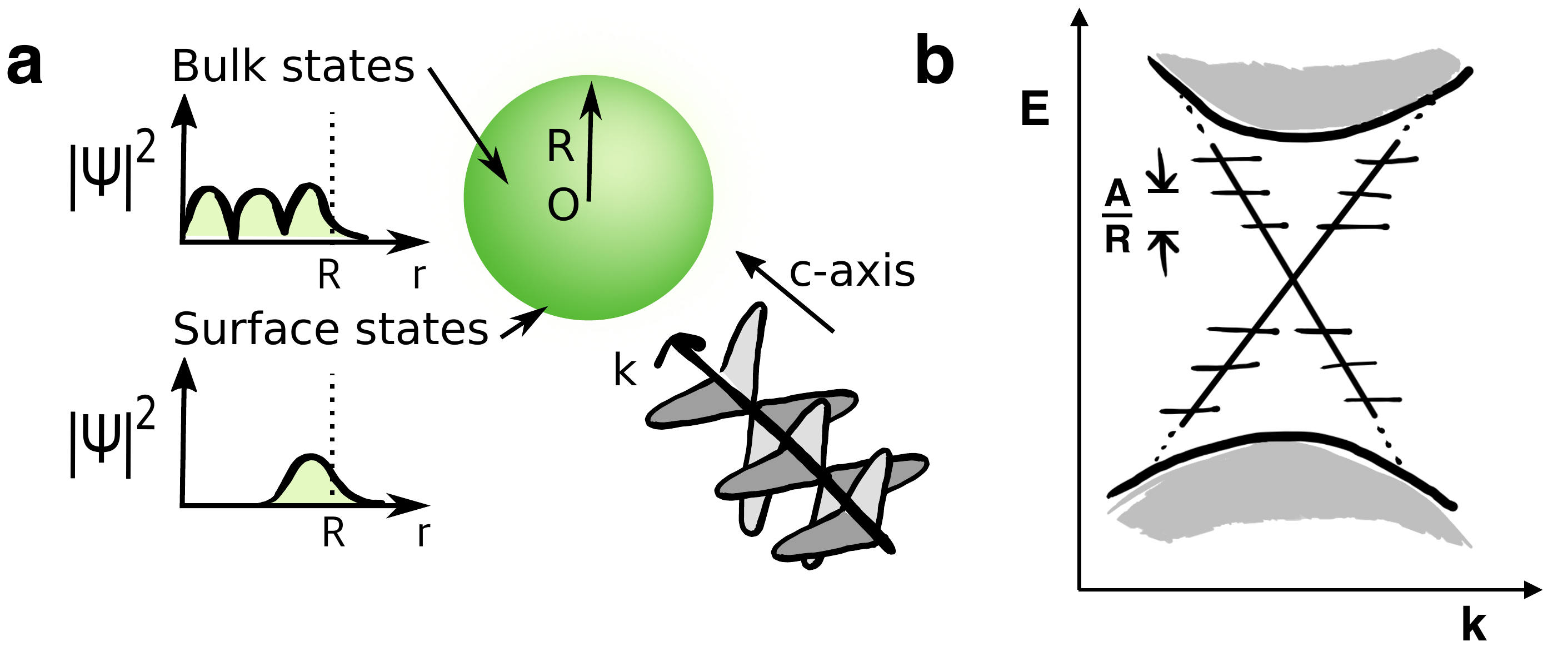}
\caption{\textbf{Topological insulator nanoparticle interacting with light} \textbf{(a)} Schematic of topological insulator nanoparticle irradiated with light along the material c-axis. Bulk states and surface states. \textbf{(b)} Discretisation of the Dirac cone, with linear spacing between states inversely proportional to the particle radius, $R$. $A$ is material dependent constant.\label{fig:tinps}}
\end{figure}

The first proposal of a 3D TI came in 2016~\cite{lu2016symmetry}, relying on a crystal symmetry (the nonsymmorphic glide reflection). The complicated nature of the structure would make it challenging to realise experimentally, whereas an all-dielectric proposal for a 3D photonic topological insulator came in 2017~\cite{slobozhanyuk2017three}, using a 3D hexagonal lattice of `meta-atoms' (dielectric disks) as displayed in FIG. \ref{fig:lit_review}g. The relative simplicity of the design and the theoretical applicability across a wide range of frequencies (including the visible) make it a likely candidate for realisation and further applications.

So far we have discussed the two types of topological photonic systems - those which explicitly break time-reversal symmetry (leading to a quantum Hall system) and those which conserve it (giving a quantum spin Hall system). When inversion symmetry is broken in a 2D honeycomb lattice, the two valleys of the band structure exhibit opposite Berry curvatures. These two valleys can be interpreted as pseudo-fermionic spins, giving rise to a time-reversal invariant effect known as the valley Hall effect. This effect has been experimentally observed~\cite{noh2018observation}, with the demonstration of counter-propagating, topological valley Hall edge states at the domain wall between two valley Hall photonic insulators of differing valley Chern number. 

\section{Topological nanophotonics} 
\label{sec:topo_nano}

We now arrive at topological nanophotonics. Some of the platforms we will describe are nanoscale versions of systems already described in section \ref{sec:topo_photonics}. However, many of these systems have fundamental size limits or are simply very difficult to engineer at the nanoscale. In some of the schemes already described, the operation frequency of the systems are fundamentally limited by the strength of the time-reversal symmetry-breaking mechanism employed, as the frequency at which the mechanism operates is too low to be used in THz platforms. We discuss some new platforms that support topological states. As described so eloquently by Toumey~\cite{let_there_be_nano}, ``nanotechnology has no single origin and spans multiple disciplines and subdisciplines all united by the same aim to control matter at the nanoscale." As such, there are many platforms on which to develop nanostructures and consequently many routes through which we can arrive at topological nanophotonics, some of which we now outline.

\begin{figure*}
\includegraphics[width=\textwidth]{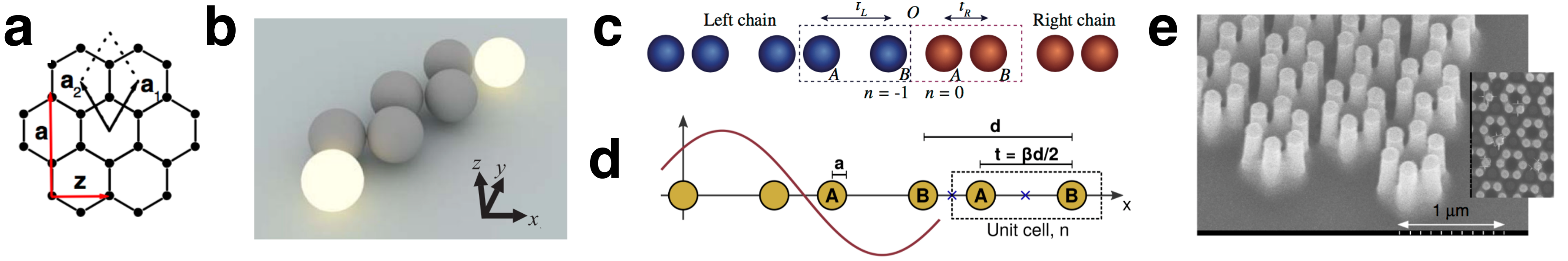}
\caption{\textbf{Proposals and realisations of topological nanophotonic platforms} \textbf{(a)} Honeycomb lattice of metallic nanoparticles supporting Dirac plasmon, reprinted figure with permission from ~\cite{han2009dirac} Copyright (2009) by the American Physical Society. \textbf{(b)} The SSH model demonstrated with dielectric nanoparticles, Reprinted figure with permission from~\cite{slobozhanyuk2015subwavelength} Copyright (2017) by the American Physical Society.
\textbf{(c)} Bipartite chain of plasmonic nanoparticles which can exhibit edge states, reproduced from ~\cite{Ling:15}. \textbf{(d)} Introducing long-range hopping with retardation and radiative damping creates a richer and more realisatic model, figure from ~\cite{doi:10.1021/acsphotonics.8b00117}. \textbf{(e)} Experimental realisation of topological states in an all-dielectric metasurface which uses far-field measurements to extract the topological invariant, reproduced from ~\cite{gorlach2018far}, CC BY 4.0. \label{fig:topo_nano}}
\end{figure*}

\subsubsection{Topological insulator nanoparticles interacting with light}
\label{subsubsec:tinps_interacting_with_light}

We saw in section \ref{subsec:tis_with_light} that electronic topological insulators will behave differently to their trivially insulating counterparts when irradiated with light. When dealing with large bulk samples of materials, the band structure of the system will exhibit a finite bulk gap, bridged by continuous, conducting topological surface states. These surface states display spin-momentum locking and as such, are extremely robust against backscattering and so unidirectional surface currents can be observed. When shrunk to the nanoscale, the surface to bulk ratio of the system becomes significant, and we can expect surface effects to have a greater impact on both the electric and optical properties of the material. 

Whilst the large TI structures of section~\ref{subsec:tis_with_light} and their interactions with light can be treated classically, for much smaller structures it is no longer efficient to treat the states of the structure classically, and we must instead treat the surface states quantum mechanically. As shown by Siroki et al~\cite{siroki2016single} in 2016, in the case of small (R$<$100nm) topological insulator nanoparticles (TINPs), the continuous Dirac cone becomes discretized due to quantum confinement effects (as illustrated in FIG. \ref{fig:topo_nano}). The discretized surface states are linearly separated by energies inversely proportional to R. If irradiated by light of commensurate frequency, single states within the Dirac cone can be excited to new localised states. This results in a new term in the absorption spectrum of the system, which (for a spherical particle of radius $R$, permittivity $\epsilon$, suspended in a background dielectric, permittivity $\epsilon_{\mathrm{out}}$) is given by 
\begin{equation}
    \sigma_{\mathrm{abs}}(\omega) = 4 \pi R^3 n_{\mathrm{out}}\frac{2\pi}{\lambda} \mathrm{Im}\left[\frac{\varepsilon +\delta_R-\varepsilon_{\mathrm{out}}}{\varepsilon +\delta_R+2\varepsilon_{\mathrm{out}}}\right],
\end{equation}
where (for Fermi level $E_F=0$) the delta contribution is given by 
\begin{equation}
    \delta_R = \frac{e^2}{6 \pi \epsilon_0}\left( \frac{1}{2A-\hbar \omega R}+\frac{1}{2A+\hbar \omega R}\right) 
\end{equation}
and $A$ is a material-dependent constant. In the absence of surface states (such as by applying a magnetic field term and thus destroying the surface states), $\delta_R$ = 0 and and we return to the usual solution of a dielectric sphere in a constant electric field.
For materials in the Bi$_2$Se$_3$ family, transitions between these topological, delocalised surface states occur within the same frequency range as a bulk phonon excitation. This results in a strong Fano resonance, referred to as the surface topological particle (SToP) mode. This is a purely quantum mechanical feature of the system, and the asymmetric profile of this resonance creates a point of zero-absorption at a particular frequency, meaning that remarkably, the excitation of a single electron occupying a topological surface state can shield the bulk from the absorption of incoming light. This mode has been theoretically predicted~\cite{siroki2016single}, and their observation in experiments is within current experimental capabilities. 
Of course, a fine tuning of the Fermi energy it is needed~\cite{Moon2018}.

The robust and discrete nature of the TINP surface states lends them to various new research paths. They may be of particular relevance in the areas of quantum optics and information, as the discrete surface states are reminiscent of an atomic scheme of energy levels, but with the additional quality of topological protection. This may make them a unique, topological type of quantum dot which could have a host of applications in areas such as topological lasing and topological quantum computing. A natural question arises now, how much big has to be a topological nanoparticle in order to preserve its topological properties, that depend on the bulk. The answer at such question was addressed using a tight-binding model and was shown particle with a diameter bigger than 5~nm behave already as a topological insulator~\cite{siroki2017protection}.


\begin{tcolorbox}[float*,colbacktitle=teal!50!white, coltitle=black,colback=teal!10!white, width=\textwidth, title={\textbf{BOX 3: Bulk-boundary correspondence and the SSH model}}]

\begin{wrapfigure}[23]{r}{0.48\textwidth}
\begin{tcolorbox}[width=0.48\textwidth, colback=white]
    \includegraphics[width=\textwidth]{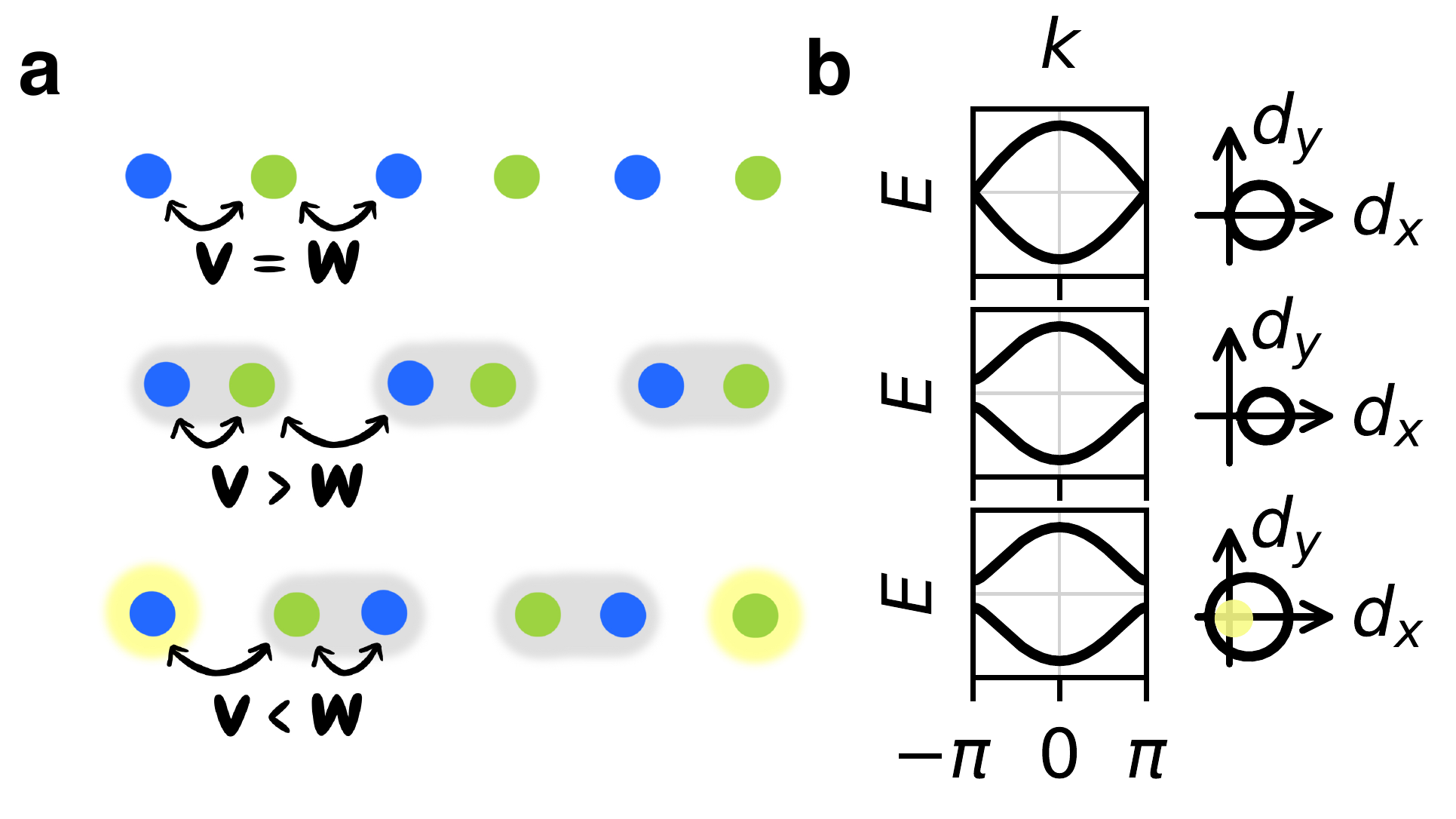}    
    FIG B3: \textbf{Bulk-boundary correspondence in the SSH model} \textbf{(a)} The three cases $v=w$, $v>w$ and $v<w$ in the dimerised limit. Subsystems $A$ and $B$ are coloured blue and green respectively.  Only the third case exhibits edge states, highlighted in yellow. \textbf{(b)} Visualization of the band structure and function $\mathbf{d}(k)$ for $-\pi<k<\pi$. The closed $\mathbf{d}(k)$ loop only encircles the origin in the last, corresponding to the existence of edge states.
\end{tcolorbox}
\end{wrapfigure}
The bulk-boundary correspondence is an important principle that tell us that the number of edge modes equals the difference in Chern numbers at that edge. An instructive way to understand it is using the topology of the SSH system first described in Box 2, we begin by studying the bulk. We first observe that the Hamiltonian of this system (equation \ref{eq:SSH_ham}) is a two band model (due to the two degrees of freedom per unit cell), and use that any two-band bulk momentum space Hamiltonian can be written in a Pauli basis, such that  $\mathcal{H} = \mathbf{d}(k)\mathbf{\sigma}$, where we introduce the basis of Pauli matrices given by 
\begin{equation}
    \begin{split}
        \sigma_x = \begin{pmatrix}0&1\\1&0\end{pmatrix},~\sigma_y = \begin{pmatrix}0&-i\\i&0\end{pmatrix},~\sigma_z = \begin{pmatrix}1&0\\0&-1\end{pmatrix}.
    \end{split}
\end{equation}
For the SSH model, the components of the vector $\mathbf{d}(k)$ are given by  $d_x(k) = v + w \mathrm{cos}(k)$, $d_y(k) = w \mathrm{sin}(k)$ and  $d_z (k) = 0$.

When plotting the band structure as $k$ goes from $0$ to $2 \pi$, we can also plot $\mathbf{d}(k)$. Due to the sublattice symmetry of the system, $d_z(k)=0$ for all $k$, and so the vector will trace out a line in the x-y plane.

We illustrate the three possible systems s for $v=w$, $v>w$ and $v<w$  in FIG B3a, and their corresponding band structures and $\mathbf{d}(k)$ plots in FIG. B3b. $\mathbf{d}(k)$ will be a closed loop due to the periodic boundary conditions of the Hamiltonian. When describing an insulator, the loop will not touch the origin (as touching the origin indicates that the gap has closed and the bands have touched, resulting in a conductor as in the case of $v=w$). For insulating phases (such as $v>w$ and $v<w$), we can count the number of times this loop winds around the origin. The number of times it goes around the origin is the bulk winding number. Note that the winding number $v$ is calculated from the bulk Hamiltonian and so is a purely bulk quantity. The bulk winding number is our first example of a topological invariant. A topological phase transition occurs at $v=w$, as the $\mathbf{d}(k)$ loop passes through the origin.

Now if we consider an open system in the dimerised limit as shown in FIG. B3A, we see that $P_A-P_B$, the net number of edge states on sublattice $A$ at the left edge is a topological invariant. For the trivial case ($v>w$), the winding number and net number of edge states are both 0. In the topological case ($v<w$) both quantities are 1. This is an illustration of the bulk-boundary correspondence, which tells us that the emergence of topological edge states is related to the topological invariants of the bulk TI system ~\cite{ryu2002topological}.
\end{tcolorbox}

\subsubsection{Systems of nanoparticles exhibiting topological phases}
Metallic nanoparticles can support localised surface plasmons, and for a system of multiple nanoparticles the near-field dipolar interactions between the plasmons cause collective plasmons. For a 2D honeycomb lattice of nanoparticles (illustrated in FIG. \ref{fig:topo_nano}a) where the  collective plasmon dispersion relation can exhibit Dirac cones, with edge states derived from the Dirac points~\cite{han2009dirac,weick2013dirac,wang2016existence}.

In 2014 it was shown that a zigzag chain of metallic nanoparticles can mimic the Kitaev wire of Majorana fermions~\cite{poddubny2014topological}. Majorana edge states are topologically protected and their manifestation in many platforms is of great interest as they are a promising candidate for topologically robus qubit states. This work shows that localised plasmons at each edge can be excited selectively, depending on the polarisation of incident light . 

For a bipartite lattice, there are various studies~\cite{Ling:15,downing2017topological,doi:10.1021/acsphotonics.8b00117,physics1010002} which demonstrate that this system can be used to construct a photonic-analogue SSH model, the electronic theory of which can be found in boxes 2 and 3 (illustrated in FIG. \ref{fig:topo_nano}c). By studying the system at the edge of the Brillouin zone it has been shown that the collective plasmons obey an effective Dirac-like Hamiltonian, and the bipartite system is governed by a non-trivial Zak phase, which predicts the topological edge states.

Recent work has considered the addition of long-range hopping into the system, with retardation and radiative damping~\cite{doi:10.1021/acsphotonics.8b00117} (shown in  FIG. \ref{fig:topo_nano}d), not only reaching towards richer and more complex physics but also creating a more realistic model, as experimental nanoscale schemes such as these will necessarily be more affected by these processes than their larger counterparts. The resulting non-Hermitian Hamiltonian displays an altered band structure, but a Zak phase and edge states which survive. 

 The above collective works demonstrate the first proposals for truly subwavelength topological states. It has also been shown that systems of dielectric nanoparticles present promising topological nanophotonic platforms. It is well known that some of the effects achieved with plasmonic nanoparticles can be reproduced using high-index dielectric particles with electric and magnetic Mie resonances, which are already used as buildings blocks in larger photonic systems (referring back to section \ref{sec:topo_photonics}). It has been shown that the SSH model can be translated into this new system of dielectric nanoparticles~\cite{slobozhanyuk2015subwavelength} (shown in FIG. \ref{fig:topo_nano}b), as well as dielectric nanodisks~\cite{kruk2017edge}. Systems of dielectric nanostructures may even present a better platform than their metallic counterparts, as they have negligible Ohmic losses, low heating and can exhibit both electric and magnetic multipolar radiation characteristics. 

\subsubsection{Graphene-based topological nanostructures}

\begin{figure}[t]
\includegraphics[width=\columnwidth]{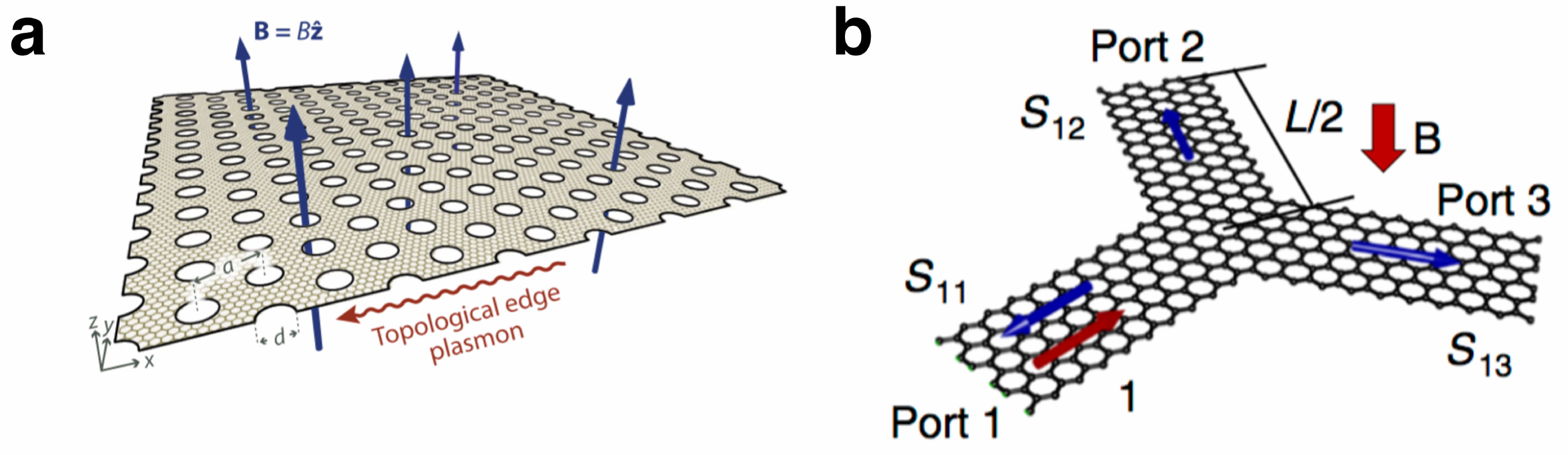}
\caption{\textbf{Graphene} 
\textbf{(a)} A doped, periodically patterned graphene sheet in an external magnetic field will exhibit unidirectional edge states. Reprinted figure with permission from~\cite{PhysRevLett.118.245301} Copyright (2017) by the American Physical Society. \textbf{(b)} More advanced schemes such as nanoribbon junctions have been presented as novel routes to topologically protected edge states, reproduced from ~\cite{pan2017topologically}, CC BY 4.0.}\label{fig:graphene}
\end{figure}

Graphene-based plasmonic crystals present another platform with which we can study topological phases in the nanoscale. Their unique plasmonic properties (tunable carrier densities, small Drude mass and long intrinsic relaxation times) allow us to study topological plasmons excitation in the THz regime. Even explicitly breaking time-reversal symmetry with a magnetic field is possible, as relatively weak magnetic fields can result in a high cyclotron frequency when combined with a small Drude mass. At finite doping of a 2D periodically patterned graphene sheet (as illustrated in FIG. \ref{fig:graphene}a), an external magnetic field will induce topologically-protected one-way edge plasmons~\cite{PhysRevLett.118.245301}. These plasmons can exist at frequencies as high as tens of THz. More complicated nanostructures can be constructed from graphene, such as nanotubes, nanocavities and even toroidal structures~\cite{kuzmin2018plasmonics}. It has also been proposed that honeycomb superlattice structures fashioned into ribbons and pierced with a magnetic field (as shown in FIG. \ref{fig:graphene}b) could present another avenue in which to produce and guide topologically protected modes within a graphene-based nanostructure~\cite{pan2017topologically}.  

\subsubsection{Topological states in metasurfaces}
\label{subsubsec:metasurfaces}
Metasurfaces are the 2D derivative of 3D metamaterials. These 2D metamaterials are made up of meta-atoms, forming a structure which is of subwavelength thickness. The spatially varying features of a metasurface can give rise to various applications, such as arbitrary wave fronts and non-linear optical effects. Their resistive loss is lower than that of 3D materials due to the absence of a bulk material, and in much the same way that 3D metamaterials can be constructed to support topological states, so can metasurfaces. Metasurfaces can be modulated both spatially and temporally (such as via external voltages or optical pumping), creating various paths towards topological phases. A recent review of progress in metasurface manufacturing and their applications has been compiled by Chang et al.~\cite{chang2018optical}. 

As with any photonic system, photonic modes can leak from the structure of a metasurface, allowing a path via which topological characteristics of the structure may be measured. As an alternative to probing topological edge states of the system, it was demonstrated in 2018 \cite{gorlach2018far} that the topological invariant of a metasurface structure can be extracted from angle-resolved spectra in the far-field (the experimental setup of which is shown in figure \ref{fig:topo_nano}e)~\cite{gorlach2018far}. 

\subsubsection{Nanophotonic topological valley Hall}

The valley Hall system shares many characteristics with the quantum spin Hall system. Broken inversion symmetry in a time-reversal symmetric 2D honeycomb lattice leads to a system in which the two valleys of the band structure exhibit opposite Chern numbers. This effect has been successful realised using a Silicon nanophotonic crystal~\cite{shalaev2018robust,barik2018robust}. Two types of crystal (with a C$_6$ symmetry and C$_3$ symmetry respectively) were fabricated, and the interface between them studied. The system with C$_6$ symmetry exhibits non-trivial topology and Dirac cones in the valleys of its bandstructure, whilst the C$_3$ structure exhibits band gaps at its valleys. Placing the two crystals together results a difference in valley Chern number across the interface, and counter-propagating edge states are necessarily localised at the interface. The study demonstrates a comparison between straight and twisted paths for the edge states, in which it is shown that the modes are very robust to harsh changes in structure such as sharp corners, and out-of-plane scattering is very low. Due to time-reversal symmetry, disorder which flips the helicity of the states can still result in backscattering however the work illustrates that on-chip fabrication of topological nano-devices is very possible, giving robust topological protection at telecommunication wavelengths.

\begin{figure*}
\includegraphics[width=\textwidth]{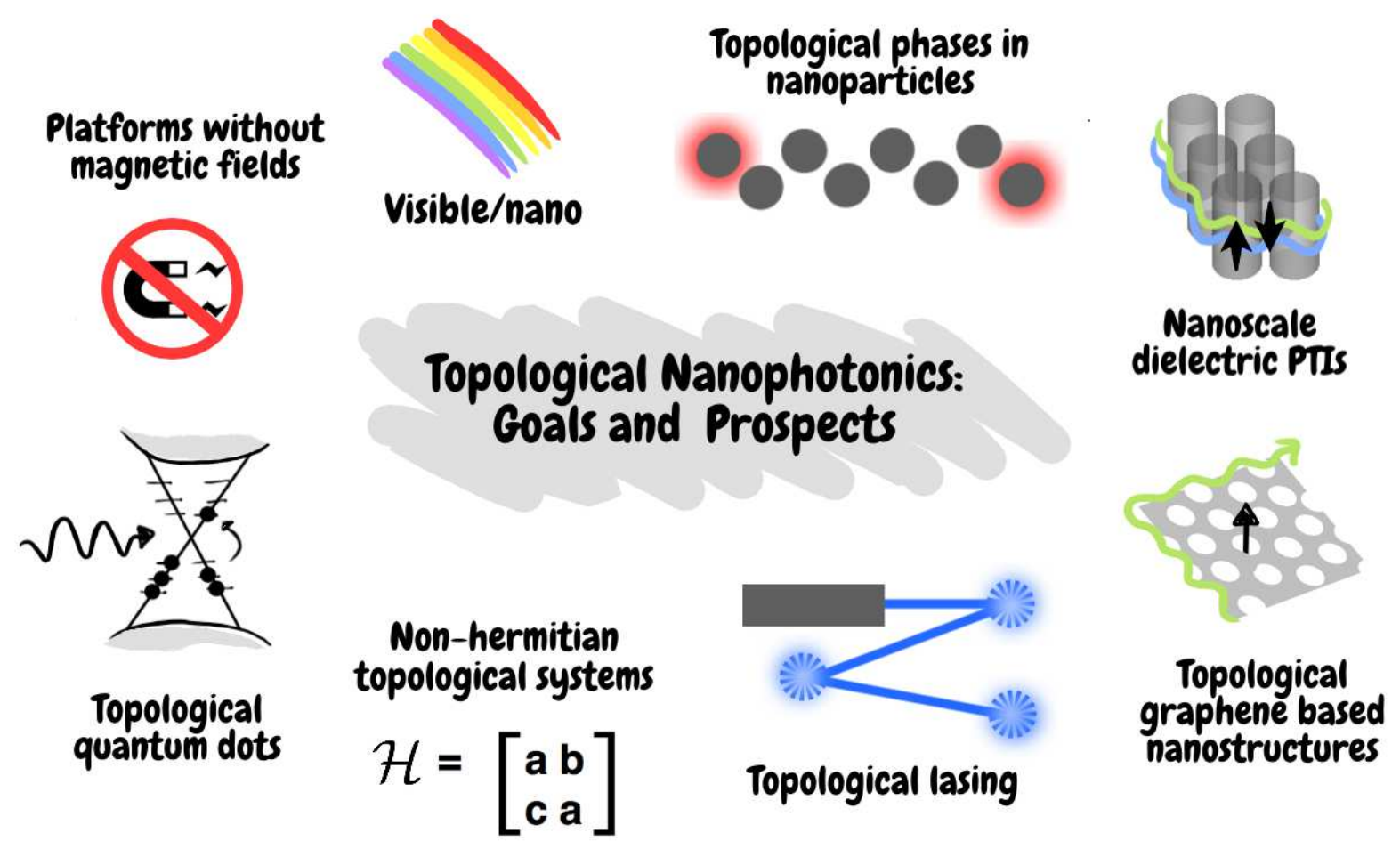}
\caption{\textbf{The goals and prospects of topological nanophotonics.} \label{fig:future}}
\end{figure*}

\section{Goals and prospects}
\label{sec:goals}

Great strides have been made in the work on topological photonics and it is now a well established and multifaceted field, but as seen in the previous section the road towards topological nanophotonics is far less travelled. 

In order to deliver the goal of topological protection of photons at the nanoscale, we look for platforms in which magnetic fields need only be of modest size or are not needed at all (as magnetic effects at the visible/nanoscale are weak). Many of the tools mastered in other frequency regimes (such as acoustic pumping or the use of metamaterials to form photonic crystals) are outside current capabilities at the nanoscale, so we look for platforms with novel ways of demonstrating and controlling topological states. 

Many of the systems of section \ref{sec:topo_nano} display Hermitian behaviour, mimicking the physics of topological electronic systems. This is because topological insulators originally emerged in the context of quantum mechanics, in which operators are Hermitian so that measurable parameters are real valued. However, the emulation of Hermitian systems is just one facet of topological photonics, which can also break Hermiticity through losses, gain and phase information. We must therefore understand how non-Hermiticity affects TIs if we hope to fully harness the power of topological protection for photonic systems.

Before their application to TIs, non-Hermitian systems have been vastly explored in the case of parity-time symmetric scenarios, where the eigenvalues of the operators are real valued despite the lack of Hermiticity. In photonics this behaviour emerges due to a balancing act between loss and gain~\cite{Feng2017nonH}. Since then parity-time (PT) symmetric systems and those with different symmetries and complex eigenvalues have been shown to exhibit topological protection, opening a door to new and exciting physics~\cite{Zeuner2015nonH,Wiemann2016nonH,Bandres2018nonH}. Researchers have predicted and observed phase transitions and edge states apparently unique to non-Hermitian systems~\cite{Leykam2017nonH,Pan2018nonH}, and are working to develop much-needed theory for non-Hermitian TIs~\cite{Kunst2018nonH,Shen2018nonH,Alvarez2018nonH}.

All Hermitian TIs are characterised by the ten-fold way, a sort of `topological insulator periodic table' associating topological behaviour with matrix symmetry classes and dimensionality~\cite{Ryu2010nonH}. The equivalent description for non-Hermitian systems is currently an open question, with recent works using the original symmetry classes of the Hermitian case or proposing the study of a larger group of symmetry classes~\cite{Gong2018nonH,Lieu2018nonH}.

The interplay between loss and gain in photonic systems already described above is not only a topic primed for scientific study, but also an opportunity for new and exciting applications. For example, the existence of robust edge states in topological systems has allowed for the breakthrough concept of topological lasing. By using topologically protected states for lasing modes, the lasing mechanism is immune to disorder under many perturbations of the system (such as local lattice deformations)~\cite{bahari2017nonreciprocal,zhao2018topological}.

Lasing from the edge states of a 1D SSH lattice has been demonstrated using polariton micropillars in the strong coupling regime~\cite{st2017lasing}. It was shown that the lasing states persist under local deformations of the lattice, and although the experiment was undertaken at low temperature (T=4K), microcavity polariton lasing experiments have been accomplished at room temperature~\cite{christopoulos2007room,kena2010room} and so this avenue has the potential to produce room-temperature lasing from topological states. 

The cavity in conventional lasers has an important role and a lot of precision and care must be taken when building a cavity. In fact, the amplification of optical modes can only happen with a properly aligned, stable cavity. An alternative approach was recently proposed~\cite{harari2018topological,bandres2018topological}, borrowing concepts from topological insulators. The main idea is a laser whose lasing mode is a topologically protected edge mode. Or in other words, we can use photonic topological crystals which present protected edge states. In this way, light which propagates in only a single direction can be amplified, despite cavity imperfections such as sharp corners or crystal defects. Such an idea has been theoretical proposed~\cite{harari2018topological} and realized~\cite{bandres2018topological} with an array of micro-ring resonators, with coupling between rings designed to follow a topological model~\cite{haldane1988model}. Allowing gain only on the edge, it is possible to enforce that the topological edge mode lases first. Such new lasing systems present interesting properties such as high slope efficiency, unidirectionality, single mode emission (even in the high gain region) aside from the extreme robustness of the mode due to topological protection.  These systems also show that non-Hermitian Hamiltonians have protected topological states, bolstering them against previous criticism. 

Despite these exciting advances, a proper theoretical background it is still missing when we deal with non-Hermitian Hamiltonians.

While they harbour great potential for robust lasing systems, topological nanophotonic systems could also be of interest in the field of topological quantum computing. Topological insulator nanoparticles exhibit a quantum dot-like structure of discrete edge states. These states can be tuned as a function of particle size, ranging from 0.003-0.03~eV and can be coupled with incoming light.  These qualities give TINPs the same functionality at a quantum dot but in the THz regime. Such topological quantum dots, if experimentally confirmed, could open new exciting paths in quantum optics at room temperature. The braiding of Majorana fermions in quantum wires~\cite{kitaev2001unpaired} has been proposed as a method of topological quantum computation, so chains of nanoparticles that can model a Kitaev chain could also follow. Any system which can robustly hold information could be of use in topological quantum computing. 

Time-modulation is a promising tool to induce topology, which has not yet been explored in nanophotonics. Photonic systems have already been used to demonstrate 4D quantum Hall physics~\cite{zilberberg2018photonic}, with the use of topological pumping. In this work, 2D arrays of evanescently coupled waveguides were used in the near-IR regime and coupled in such a way that momenta associated with two synthetic dimensions were sampled and a 2D topological pump was realized. The resulting band structure of the light has a second Chern number associated with a 4D symmetry, and the photon pumping in this system is analogous to charge pumping in an electronic system. Pumping in the visible range is difficult due to the high operating frequency, but possibilities at the nanoscale do exist - for instance Graphene can be modulated at hundred of GHz, whilst hosting plasmons in the THz. The interplay of non-Hermiticity and topological gaps associated with non-zero second Chern numbers has yet to be explored, and nanophotonic systems provide a possible platform from which to study this type of interplay and the related physics of topological systems. 

In order to obtain strong interaction between light and matter, an important goal will be to move the topological insulator properties to higher frequencies as visible an near-UV.

The plethora of possibilities and new paradigms available in the topic of topological nanophotonics (as illustrated in FIG. \ref{fig:future}) make it an exciting field to study and strong theoretical and experimental challenge. With technical feats of nanofabrication improving steadily, the potential for topological protection and precise control of photons at the nanoscale is extensive and there is much yet which can be accomplished in this new and rapidly developing field. 

\section*{Acknowledgements}
MSR and SJP would like to acknowledge their studentships from the Centre for Doctoral Training on Theory and Simulation of Materials at Imperial College London funded by EPSRC Grant No. EP/L015579/1. SRP acknowledges funding from EPSRC. XX is supported by a Lee family scholarship. PAH acknowledges The Gordon and Betty Moore Foundation. VG acknowledges the Spanish Ministerio de Economia y Competitividad for financial support
through the grant NANOTOPO (FIS2017-91413-EXP), and also Consejo Superior de Investigaciones Científicas (INTRAMURALES 201750I039).

\bibliographystyle{unsrt}

\end{document}